\documentclass[10pt,letterpaper]{article}
\usepackage{opex3}

\begin{document}

\title{Resonant-tunnelling-assisted crossing for subwavelength plasmonic slot waveguides}

\author{Sanshui Xiao and Niels Asger Mortensen}

\address{DTU Fotonik - Department of Optics Engineering, \\Technical
University of Denmark, DK-2800 Kongens Lyngby, Denmark.}

\email{saxi@fotonik.dtu.dk} 



\begin{abstract}
We theoretically investigate the properties of crossing for two
perpendicular subwavelength plasmonic slot waveguides. We show
that, when encountering nano intersection, the crosstalk for the
direct crossing is around $25\%$, almost same as throughout. In
terms of symmetry considerations and resonant-tunnelling effect,
we design compact cavity-based structures. Our results show that
the crosstalk is eliminated and the throughput reaches the unity
on resonance. Simulations results are in agreement with those from
the coupled-model theory. When taking into account of the material
loss, due to the unchanged symmetry properties of the modes, the
crosstalk is still suppressed. Our results may open a way to
construct nanoscale crossings for high-density nanoplasmonic
integration circuits.
\end{abstract}



\section{Introduction}
It is of great interest for guiding light at deep subwavelength
scales in optoelectronics, partly because it may enable
ultra-density integration of optoelectronic circuits. Conventional
dielectric waveguides cannot restrict the spatial localization of
optical energy beyond the $\lambda_0/2n$ limit
\cite{Lipson:2005,Bogaerts:2005,Tsuchizawa:2005,Takahara:1997},
where $\lambda_0$ is the free space photon wavelength and $n$ is
the refractive index of the waveguide. Surface plasmon polaritons
(SPPs) waveguides, which utilize the fact that light can be
confined at metal-dielectric interface, have shown the potential
to guide and manipulate light at deep subwavelength
scales\cite{Ebbesen:2008,Maier:2003,Tanaka:2003,Veronis:2005,Liu:2005,Pile:2005,Bozhevolnyi:2006,Zia:2004}.
The prospect of integration has motivated significantly recent
activities in exploring plasmonic waveguide structures. In
constructing highly dense integration of optoelectronic circuits,
the ability to intersect waveguides is crucial owing to the desire
for complex systems involving multiple waveguides. Usually,
waveguide crossings require low intersection loss, low crosstalk,
and compact dimensions. Several designs have been proposed for
low-loss, low-crosstalk crossing of silicon-on-insulator
nanophotonic waveguides
\cite{Bogaerts:2007,Fukazawa:2004,Chen:2006}. However, to our
knowledge, there are few studies about the waveguide crossings for
SPP waveguides. Note that previously much attention has been
focused on realizing the surface plasmon waveguide with long
propagation length \cite{Sarid:1981,Berini:2000,Jung:2007}. In
this paper, we analysis the intersection loss of nanoplasmonic
waveguide and design compact intersections with no crosstalk,
based on resonant tunnelling effect. Results may open a way to
construct nanoscale crossings for high-density nanoplasmonic
integration circuits. Except for the dispersion of the SPP
waveguide, the calculations mentioned below are performed by the
finite-element method (FEM) in frequency domain.

\section{Dispersion of surface plasmon polariton waveguide}
\begin{figure}[htbp]
\centering\includegraphics[width=6cm]{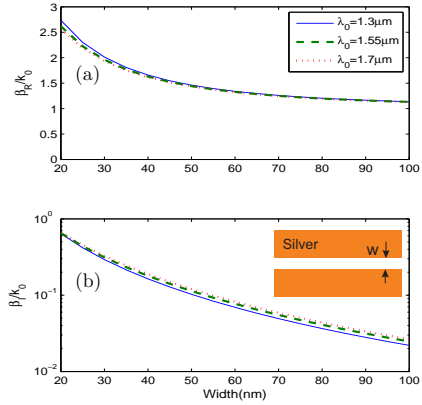}
\caption{Dependence of complex propagation constants of SPPs in a
2D silver-air-silver plasmonic waveguide on the width ($w$) of the
waveguide and the working wavelength $\lambda_0$. }
\label{Dispersion}
\end{figure}

Consider a subwavelength  metal-dielectric-metal (MDM) constructed
two-dimensional (2D) plasmonic waveguide. The propagation constant
$\beta (=\beta_R+j\beta_I)$ of surface plasmon polaritons can be
obtained by solving the dispersion equation:
\begin{eqnarray}
\frac{\varepsilon_1p}{\varepsilon_mk}=\frac{1-exp(kw)}{1+exp(kw)},
\end{eqnarray}
where $k$ and $p$, the functions of $\beta$, are the wave numbers
of SPPs in dielectric and metal, respectively, $\varepsilon_1$ and
$\varepsilon_m$ are the dielectric constants of the medium in
guide region and metals, respectively, and $w$ is the width of the
waveguide. For such a 2D plasmonic waveguide, the fundamental
transverse magnetic mode ($TM_0$) always exists even when the
width is close to zero, while other high-order modes have a cutoff
width. To satisfy single-mode condition, the width of the
plasmonic waveguide should be smaller than
$\lambda_0Atan(\sqrt{-Re(\varepsilon_m)/\varepsilon_1})/(\pi\sqrt{\varepsilon_1})$
, where $Atan$ is the mathematical arctangent function. For
instance, the maximum width of the single-mode condition for the
silver-air-silver SPP waveguide is about 720nm for
$\lambda_0=1.55\mu m$. Figure \ref{Dispersion} shows the
dependence of $\beta/k_0$ of the fundamental SPP mode in 2D
silver-air-silver waveguide on the width ($w$) of the waveguide
and working wavelength $\lambda_0$ of light in free space, where
$k_0=2\pi/\lambda_0$. From Fig. \ref{Dispersion}, one can see that
the effective refractive index ($n_{eff}=\beta_R/k_0$) of SPP
$TM_0$ mode is always larger than that of the dielectric medium,
i. e., $n_{eff}>\sqrt{\varepsilon_1}$. The loss arisen from the
intrinsic loss of the metal increases when shrinking the width of
the plasmonic waveguide. In this paper, the waveguide width is
chosen to be much smaller than the waveguide to have the
single-mode property of the SPP waveguide.

\section{Direct crossing for two perpendicular plasmonic waveguides}
When considering the waveguide crossing, the confinement in the
direction perpendicular to the wave propagation is lost near the
crossing region, thus causing diffraction of the light. The
diffraction strongly depends on size and the index-contrast $\eta$
of the waveguide. The loss for the waveguide crossing of
low-index-contrast waveguides is negligible, while the mode
diffracts dramatically for the nano-size of high-index-contrast
waveguides. Previous studies show that the high-index-contrast
systems, such as silicon-on-insulator nanowires with $\eta$ of
2.34 ($\eta={n_{silicon}}/{n_{silica}}$), have a large
intersection loss for the direct waveguide crossing. The
silver-air-silver plasmonic waveguide studied here is a size of a
few hundred nanometers, having a quite large index-contrast of
~9.3($\eta={n_{silver}}/{n_{air}}$) for the working wavelength
1.55 $\mu m$. One believes that the SPP mode will be significantly
diffracted when passing through a waveguide crossing with nano
size.
\begin{figure}[htbp]
\centering\includegraphics[width=12cm]{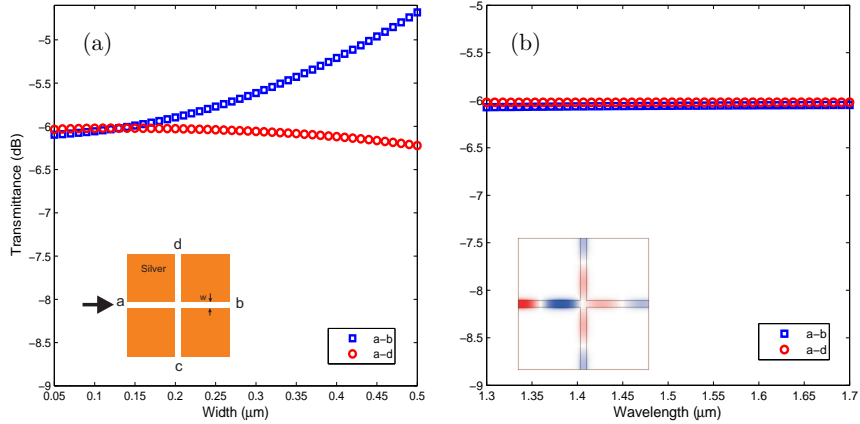} \caption{(a) The
straightforward transmittance (squares) and crosstalk (circles)
for the standard direct crossing as a function of the width ($w$)
of the silver-air-silver plasmonic waveguide for $\lambda_0$=1.55
$\mu m$. The inset shows its corresponding structure. (b)
Transmittance spectra for the straightforward (a-b) and crosstalk
(a-d) lines for $w$=100 $nm$. The inset shows the profile of a
steady-state magnetic field at the wavelength 1.55$\mu m$.}
\label{DirectCrossing}
\end{figure}
To clearly illustrate it, we first analysis the behavior of
intersection loss for the direct waveguide crossing of
nanoplasmonic waveguides, as shown in the inset of Fig.
\ref{DirectCrossing}(a). Suppose an optical beam incident from
port a. Obviously the beam diffracts when encountering the
crossing region. The fraction power transmission are shown in Fig.
\ref{DirectCrossing}. For simplicity, here we assume that the
metal is lossless, i.e, neglecting the imaginary part
$Im(\varepsilon_m)$ of metal dielectric permittivity. Fig.
\ref{DirectCrossing} (a) shows the fraction power transmission as
a function of the width of the plasmonic silver-air-silver
waveguide for the working wavelength 1.55 $\mu m$, where $w$
denotes the width of the waveguide. The squares and circles
represent the straightforward transmittance (a-b) and crosstalk
(a-d), respectively. The straightforward transmittance drops down
when the width of the plasmonic waveguide decreases from 500 $nm$
to 50 $nm$, and the crosstalk slightly increases when decreasing
the waveguide's width. For the case of $w$=200 $nm$, the
straightforward throughout is about -5.89dB ($25.73\%$) and
crosstalk is about -6.03dB ($24.96\%$). From Fig.
\ref{DirectCrossing} (a), one concludes this kind of direct
waveguide crossing has low transmission and high crosstalk. It is
also interesting to note that, when encountering nano
intersection, the throughout is around -6.00dB ($25\%$), almost
same as crosstalk. The transmittance spectra, for the case of
$w$=100 $nm$,  with straightforward line (a-b) and crosstalk line
(a-d) are shown in Fig. \ref{DirectCrossing} (b). In all
calculations mentioned in this paper, frequency-dependent
dielectric function of the Silver is described by the lossy Drude
model
$\varepsilon(\omega)=\varepsilon_\infty-(\varepsilon_0-\varepsilon_\infty)\omega_p^2/(\omega^2+2i\omega\nu_c)$,
where $\varepsilon_\infty$/$\varepsilon_0$ is the relative
permittivity at infinite/zero frequency, $\omega_p$ is the plasma
frequency, and $\nu_c$ is the collision frequency. We choose
$\varepsilon_\infty=4.017$, $\varepsilon_0=4.688$,
$\omega_p=1.419\times 10^{16}rad/s$ and $\nu_c=1.117\times
10^{14}rad/s$ for the Drude model, which fits the experimental
data \cite{Palikbook} quite well. For the wavelength of our
interests, the throughout keeps the same value of -6.00dB
($25\%$), as well as the crosstalk. The inset of Fig.
\ref{DirectCrossing} (b) shows the profile of a steady-state
magnetic field at the working wavelength 1.55 $\mu m$, which
illustrates that the throughout is almost same as the crosstalk.

\section{Resonant-tunnelling assisted transmittance for subwavelength plasmonic slot waveguide}
\begin{figure}[htbp]
\centering\includegraphics[width=13cm]{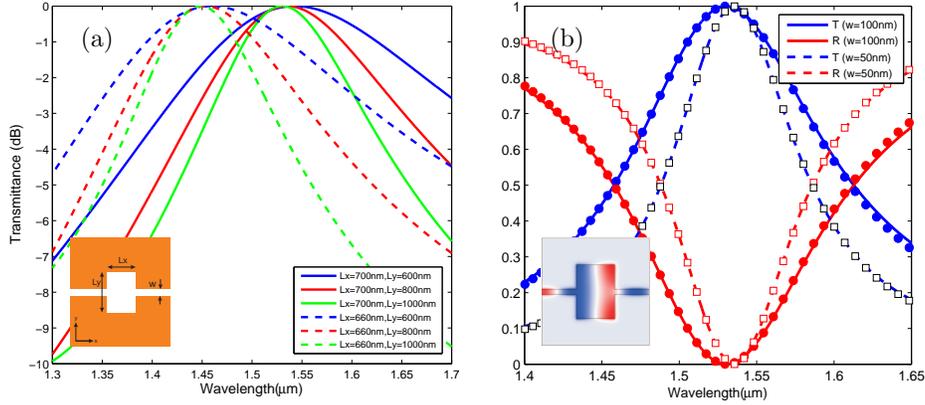} \caption{(a)
Transmission spectra of the device [shown in the inset] with
different side length of the cavity, where the width of the
plasmonic waveguide is fixed as 100 $nm$. The solid and dashed
lines represent the results for $Lx$=700 $nm$ and 660 $nm$,
respectively. (b) Transmission $T$ and reflection $R$ of the
device ($Lx$=700 $nm$, $Ly$=1000 $nm$) as a function of the
wavelength for $w$=50 $nm$ and 100 $nm$, respectively. The solid
and dashed lines represent the results from the FEM method, and
the open squares and solid circles are obtained from the
coupled-mode theory. The inset shows the profile of a steady-state
magnetic field at the resonant frequency for $w$=100 $nm$, which
illustrates the complete transmission on resonance.}
\label{Onewaveguide}
\end{figure}
To decrease the crosstalk and enhance the straightforward
transmittance of the crossing waveguide intersection, one can aim
to reduce the diffraction in the crossing region. In order to
suppress effectively the diffraction, generally an intersection
region with a large size of numbers of wavelength is used, which
results in low packing density of optical circuits. In this paper,
we utilize the well-known effect of resonant tunnelling through a
cavity to eliminate the crosstalk and increase the straightforward
transmittance. Firstly we consider a system consisting of a
subwavelength silver-air-silver plasmonic waveguide coupled to a
rectangular cavity, which supports a resonant mode of frequency
$\omega_0$. For such a system, the transmission can be described
by the resonant tunnelling effect, and one can use the
coupled-mode theory \cite{Hausbook} to evaluate the power
transmission $T$ and reflection $R$ on resonance by
\begin{eqnarray}
&T = \frac{\left(\frac{\displaystyle 1}{\displaystyle \tau_e}\right)^2}
{\left(\frac{\displaystyle 1}{\displaystyle \tau_0}+\frac{\displaystyle 1}{\displaystyle \tau_e}\right)^2}, \label{coupledmode}\\
&R = \frac{\left(\frac{\displaystyle 1}{\displaystyle
\tau_0}\right)^2}{\left(\frac{\displaystyle 1}{\displaystyle
\tau_0}+\frac{\displaystyle 1}{\displaystyle \tau_e}\right)^2},
\label{coupledmode2}
\end{eqnarray}
where $1/\tau_0$ is the decay rate due to the internal loss in the
cavity and $1/\tau_e$ is the decay rate of the field in the cavity
due to the power escape through the waveguide. From the above
equations, one can see the direct relation between the
transmittance/reflection and the ratio $\tau_0/\tau_e$ on
resonance. If there is no internal loss in the cavity
($1/\tau_0=0$), the incident wave is completely transmitted and
the spectral width of the resonance is determined by the strength
of the coupling between the waveguide and the cavity ($1/\tau_e$).
Consider a resonant system, shown in the inset of Fig.
\ref{Onewaveguide}(a). To keep structural symmetry, the center of
the cavity is placed at the center of the plasmonic waveguide and
the side length of the rectangular cavity is denoted by $Lx/Ly$.
Assuming that the metal is lossless, the transmission $T$
coefficients of the device [inset of Fig. \ref{Onewaveguide}(a)]
with different side length of the cavity are shown in Fig.
\ref{Onewaveguide}(a), where the width of the plasmonic waveguide
$w$ is fixed as 100 $nm$. From Fig. \ref{Onewaveguide}(a), we
observe that the resonant frequency of the cavity strongly depends
on the side length of the cavity in $x$ direction $Lx$, while
slightly shifts when varying $Ly$. We also find out that, for a
fixed value of $Lx$, the quality factor
$Q_{total}$($1/Q_{total}=1/Q_{coupling}+1/Q_{intrinsic}$) of the
resonant system increases when enlarging $Ly$. $Q_{total}$ is
around 5 when $Lx$=700 $nm$, $Ly$=600 $nm$, and increases to 10
for $Lx$=700 $nm$, $Ly$=1000 $nm$. Fig. \ref{Onewaveguide}(b)
shows the transmission $T$ and reflection $R$ of the device
($Lx$=700 $nm$, $Ly$=1000 $nm$) as a function of the wavelength
for $w$=50 $nm$ and $w$=100 $nm$, respectively. The solid and
dashed lines in Fig. \ref{Onewaveguide}(b) represent the results
from the FEM method, and the open squares and solid circles in
Fig. \ref{Onewaveguide}(b) are obtained from the coupled-mode
theory. Results from coupled-mode theory are in agreement with
those from the FEM method. Since the metal is assumed to be
lossless, there is no internal loss in the cavity and there is,
therefore, complete transmission on resonance, as seen in Fig.
\ref{Onewaveguide}(a) and Fig. \ref{Onewaveguide}(b). In this
coupling system, the coupling strength can be tuned by the width
$w$ of the plasmonic waveguide. Decreased the width results in a
weaker coupling and, therefore, higher quality factor and narrower
spectral width of the resonance. For the case of $w$=100 $nm$,
$Q_{total}$ is about 10 and becomes 15 for $w$=50 $nm$. For the
lossless case, because of infinitive $Q_{intrinsic}$, $Q_{total}$
is solely determined by the coupling strength $Q_{coupling}$,
$i.e.$, $Q_{total}=Q_{coupling}$. From Fig. \ref{Onewaveguide}(b),
we also observe that the resonant frequency of the cavity slightly
shifts when $w$ is varied. The inset of Fig. \ref{Onewaveguide}
(b) shows the profile of a steady-state magnetic field at the
resonant frequency for $w$=100 $nm$, which illustrates the
complete transmission on resonance. From the mode distribution in
the rectangular cavity[inset of Fig. \ref{Onewaveguide}(b)], we
can explain why the resonant frequency of the cavity is strongly
dependent to $Lx$, while almost independent to $Ly$, as mentioned
above.
\begin{figure}[htbp]
\centering\includegraphics[width=8cm]{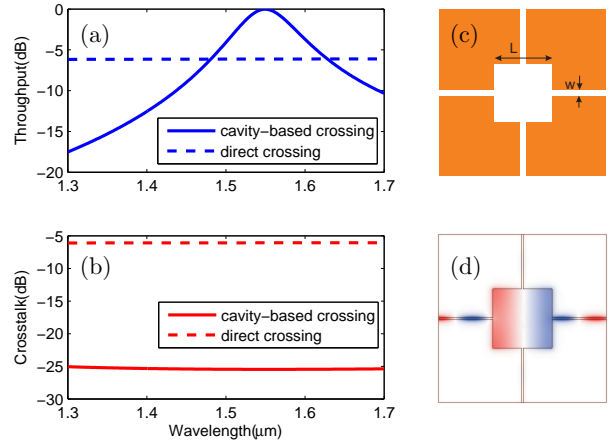} \caption{(a)-(b)
Transmission spectra of the device with direct crossing (the
dashed lines) and cavity-based crossing (the solid lines). (c)
Intersection of two-dimensional plasmonic waveguides with a
resonant cavity. (d) Profile of a steady-state magnetic field at
the resonant wavelength 1.55$\mu m$ for $w$=20 $nm$ and $L$=700
$nm$.} \label{Crosstalk}
\end{figure}

Next let us consider coupling of the four branches of the
perpendicular intersection for two plasmonic waveguides in terms
of a resonant cavity at the center, as shown in Fig.
\ref{Crosstalk} (c). Based on symmetry considerations, when the
resonant mode that is excited from the input port can be prevented
from decaying into the transverse ports, the crosstalk can be
prohibited and the system reduces to the resonant tunnelling
phenomenon through a cavity. To achieve it, there are general
criteria for perpendicular intersection of two waveguides, as
mentioned in detail in Ref. \cite{Johnson:1998}. To achieve it,
the following conditions should be satisfied: (1) Waveguide:
single-mode with mirror symmetry plane through its axis and
perpendicular to the other one; (2) Cavity: symmetric with respect
to the mirror planes of both waveguides; (3) Cavity modes: only
two degenerate modes with different symmetry with respect to
waveguide's mirror plane. When these requirements are satisfied,
due to its orthogonality to the mode in the other waveguide, each
resonant state can couple solely to the mode in just one
waveguide, thus the crosstalk will be eliminated. From the mode
profile [inset of Fig. \ref{Onewaveguide} (b)], one can easily see
that the resonant mode supported by the rectangular cavity is even
with respect to one waveguide's mirror plane and odd with respect
to the other and that there is only one resonant mode in the
wavelength range of interest in Fig. \ref{Onewaveguide}. In terms
of the general criteria mentioned above, the crosstalk for the
perpendicular intersection [Fig. \ref{Crosstalk} (c)] will be
possibly inhibited when introducing a square cavity in the
intersection. One also note that for the 2D case the bulk material
(Silver) can prevent any radiation losses. Following this idea, we
design quite a simple crossing intersection with a square resonant
cavity, as shown in Fig. \ref{Crosstalk}(c). Figs. \ref{Crosstalk}
(a) and \ref{Crosstalk} (b) show the transmission spectra of the
devices with direct crossing and cavity-based crossing (the solid
line). Compared with the result for direct crossing (the dashed
line), the throughput (solid lines) in Fig. \ref{Crosstalk} (a) is
really enhanced due to the resonant-tunnelling effect and reached
the unity on resonance. The crosstalk (the solid line) for
cavity-assisted crossing, shown in Fig. \ref{Crosstalk}(b), is
almost prohibited relative to unmodified crossings. The crosstalk
is close to zero in the whole frequency range of interest, which
can be naturally understood by the general criteria mentioned
above. Fig. \ref{Crosstalk} (d) shows of a steady-state magnetic
field at the resonant frequency for $w$=20 $nm$, which illustrates
the energy is fully transmitted forward through the crossing
section. We also note that the size of the intersection is quite
compact, which is vital for high-density integration.

\begin{figure}[htbp]
\centering\includegraphics[width=6.5cm]{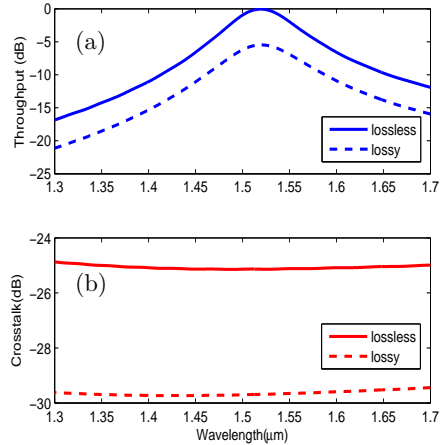} \caption{(a)
Throughput of the device [see Fig. \ref{Crosstalk} (c)] with
ignoring the material loss (the solid line) and considering the
loss (the dashed line). (b) Crosstalk of the device with ignoring
the material loss (the solid line) and considering the loss (the
dashed line). } \label{Loss}
\end{figure}

All calculations mentioned above are performed when ignoring the
material loss.  Note that the metal, Silver, is always a lossy
material especially in the visible and infrared frequency region.
This kind of loss usually limits the optical performance of the
plasmonic devices. For the device studied here, the loss will
result in large propagation loss of the plasmonic waveguide and
low quality factor of the unloaded resonant plasmonic cavity.
However, when we take into account of it, the symmetric properties
of cavity modes will keep unchanged, as well as that for the
waveguide mode. We believe that the general criteria for
eliminating the crosstalk of the perpendicular intersection are
still satisfied, i. e.,  the crosstalk can also be inhibited when
taking into account of the material loss. From Eqs.
(\ref{coupledmode})-(\ref{coupledmode2}), one knows that the
spectrum will not reach unity on resonance. This is because the
factor $\frac{\displaystyle 1}{\displaystyle \tau_o}$
significantly decreases when considering the material loss and
becomes comparable to $\frac{\displaystyle 1}{\displaystyle
\tau_e}$. Here, we recalculate the transmission spectra for the
perpendicular intersection [see Fig. \ref{Crosstalk} (c)] without
ignoring the material loss, and results are shown in Fig.
\ref{Loss}. Figs. \ref{Loss} (a) and \ref{Loss} (b) show the
transmission spectra of the devices [see Fig. \ref{Crosstalk} (c),
$L$=680 $nm$, $w$=20 $nm$]. The solid and dashed lines represent
the results when ignoring the material loss (the solid line) and
considering the loss (the dashed line), respectively. When we take
into account of the loss, the throughput shown in Fig. \ref{Loss}
(a) is only about -5.47dB on resonance, which is arisen from the
propagation loss of the waveguide mode and the reflection by the
plasmonic cavity. At telecommunication windows, around 1.5 $\mu
m$, the SPP propagation length is greater than the total length of
the circuit and can reach values close to 1 $mm$. The propagation
loss can be possibly miniaturized for each plasmon device. Apart
from the propagation loss, the intersection loss for the device
[see Fig. \ref{Crosstalk} (c)]is about -0.27dB on resonance, which
is limited by the value of $Q_{intrinsic}/Q_{coupling}$. For the
lossless case, the $Q_{total}$ is around 20 and becomes 15 when
considering the loss. $Q_{intrinsic}$ of the unloaded cavity with
the lossless case is much larger than $Q_{coupling}$, thus we can
obtain $Q_{coupling}=Q_{total}=20$. Due to the metal loss,
$Q_{intrinsic}$ strongly decreases. Assuming that the coupling
strength are the same for both cases, one can obtain that
$Q_{intrinsic}$ becomes 60 for the lossy case. In order to improve
the transmittance property, what we can do is to increase the
value of $Q_{intrinsic}/Q_{coupling}$. Recently, people use the
gain material to annul the effect of material loss, thus improving
the optical performance of loss-limited plasmonic
devices\cite{Maier:2006,Nezhad:2006}. From Fig. \ref{Loss}(b), one
can also observe that the crosstalk is also prohibited without
ignoring the loss. The simulation results agree very well with
what we expect in the analysis mentioned above.

\section{Summary}
In this paper, we analysis the intersection loss for two
perpendicular plasmonic waveguides. For the direct crossing, when
encountering nano intersection the throughout is around $25\%$,
almost same as crosstalk. Using a general recipe for elimination
of crosstalk, we design simple cavity-based structures to enhance
the throughput and eliminate the crosstalk. The size of the
intersection is quite compact, which is vital for high-density
integration. Numerical results are calculated by FEM in frequency
domain, which agree well with those from the coupled-model theory.
Without considering the material loss, the throughput reaches the
unity on resonance and the crosstalk is suppressed, close to zero.
Results can be explained in terms of symmetry considerations and
resonant tunnelling effects. For the lossy case, due to the
unchanged symmetry properties of the cavity modes and waveguide
mode, the crosstalk is also suppressed and the throughput never
reaches the unity on resonance. Apart from the propagation loss,
the intersection loss for the device is about -0.27dB on
resonance. Our results may open a way to construct nanoscale
crossings for high-density nanoplasmonic integration circuits.

\section*{Acknowledgments}
This work is financially supported by the Danish Research Council
for Technology and Production Sciences (grants no: 274-07-0379,
274-07-0080) as well as the Danish Council for Strategic Research
through the Strategic Program for Young Researchers (grant no:
2117-05-0037).


\begin{thebibliography}{10}
\newcommand{\enquote}[1]{``#1''}
\expandafter\ifx\csname url\endcsname\relax
  \def\url#1{\texttt{#1}}\fi
\expandafter\ifx\csname
urlprefix\endcsname\relax\def\urlprefix{URL }\fi
\providecommand{\eprint}[2][]{\url{#2}}

\bibitem{Lipson:2005}
M.~Lipson, \enquote{Guiding, modulating, and emitting light on
Silicon -
  Challeges and opportunities,} J. Lightwave Technol. \textbf{23}, 4222--4238 (2005).

\bibitem{Bogaerts:2005}
W.~Bogaerts, R.~Baets, P.~Dumon, V.~Wiaux, S.~Beckx, D.~Taillaert,
  B.~Luyssaert, J.~Van~Campenhout, P.~Bienstman, and D.~Van~Thourhout,
  \enquote{Nanophotonic waveguides in silicon-on-insulator fabricated with CMOS
  technology,} J. Lightwave Technol. \textbf{23}, 401--412 (2005).

\bibitem{Tsuchizawa:2005}
T.~Tsuchizawa, K.~Yamada, H.~Fukuda, T.~Watanabe, J.~Takahashi,
M.~Takahashi,
  T.~Shoji, E.~Tamechika, S.~Itabashi, and H.~Morita, \enquote{Microphotonics
  Devices Based on Silicon Microfabrication Technology,} IEEE J. Sel. Top.
  Quantum Electron \textbf{11}, 232--240 (2005).

\bibitem{Takahara:1997}
J.~Takahara, S.~Yamagishi, H.~Taki, A.~Morimoto, and T.~Kobayashi,
  \enquote{Guiding of a one-dimensional optical beam with nanometer diameter,}
  Opt. Lett. \textbf{22}, 475--477 (1997).

\bibitem{Ebbesen:2008}
T.~W. Ebbesen, C. Genet, and S. I. Bozhebolnyi,
\enquote{Surface-plasmon circuitry,} Phys. Today \textbf{61},
44--50 (2008).

\bibitem{Maier:2003}
S.~A. Maier, P.~G. Kik, H.~A. Atwater, S.~Meltzer, E.~Harel, B.~E.
Koel, and
  A.~A.~G. Requicha, \enquote{Local detection of electromagnetic energy
  transport below the diffraction limit in metal nanoparticle plasmon
  waveguides,} Nat. Mater. \textbf{2}, 229--232 (2003).

\bibitem{Tanaka:2003}
K.~Tanaka and M.~Tanaka, \enquote{Simulations of nanometric
optical circuits
  based on surface plasmon polariton gap waveguide,} Opt. Lett.
  \textbf{82}, 1158--1160 (2003).

\bibitem{Veronis:2005}
G.~Veronis and S.~Fan, \enquote{Guided subwavelength plasmonic
mode supported
  by a slot in a thin metal film,} Opt. Lett. \textbf{30}, 3359--3361
  (2005).

\bibitem{Liu:2005}
L.~Liu, Z.~Han, and S.~He, \enquote{Novel surface plasmon
waveguide for high
  integration,} Opt. Express \textbf{13}, 6645--6650 (2005).

\bibitem{Pile:2005}
D.~F.~P. Pile, T.~Ogawa, D.~K. Gramotven, Y.~Matsuzaki, K.~C.
Vernon,
  T.~Yamaguchi, K.~Okamoto, M.~Haraguchi, and M.~Fukui,
  \enquote{Two-dimensionallly localized modes of a nanoscale gap plasmon
  waveguide,} Appl. Phys. Lett. \textbf{87}, 261114 (2005).

\bibitem{Bozhevolnyi:2006}
S.~I. Bozhevolnyi, V.~S. Volkov, E.~Devaux, J.~Y. Laluet, and
T.~W. Ebbesen,
  \enquote{Channel plasmon subwavelength waveguide components including
  interferometers and ring resonators,} Nature \textbf{440}, 508--511
  (2006).

\bibitem{Zia:2004}
R.~Zia, M.~D. Selker, P.~B. Catrysse, and M.~L. Brongersma,
\enquote{Geometries
  and materials for subwavelength surface plasmon modes,} J. Opt. Soc. Am. A
  \textbf{21}, 2442--2446 (2004).

\bibitem{Bogaerts:2007}
W.~Bogaerts, P.~Dumon, D.~V. Thourhout, and R.~Baets,
\enquote{Low-loss,
  low-cross-talk crossings for silicon-on-insulator nanophotonic waveguides,}
  Opt. Lett. \textbf{32}, 2801--2803 (2007).

\bibitem{Fukazawa:2004}
T.~Fukazawa, T.~Hirano, F.~Ohno, and T.~Baba, \enquote{Low loss
intersection of
  Si photonic wire waveguides,} Jpn. J. Appl. Phys. Part 1 \textbf{43}, 646--647 (2004).

\bibitem{Chen:2006}
H.~Chen and A.~W. Poon, \enquote{Low-loss
multimode-interference-based
  crossings for silicon wire waveguides,} IEEE Photon. Technol. Lett.
  \textbf{18}, 2260--2262 (2006).

\bibitem{Sarid:1981}
D. ~Sarid, \enquote{Long-Range surface-plasma waves on very thin
metal films,} Phys. Rev. Lett.,
  \textbf{47}, 1927--1930 (1981).

\bibitem{Berini:2000}
P. ~Berini, \enquote{Plasmon-polariton waves guided by thin lossy
metal films of finite width: Bound modes of symmetric structures,}
Phys. Rev. B,
  \textbf{61}, 10484--10503 (2000).

\bibitem{Jung:2007}
J. ~Jung, T. ~Sondergaard, and S. I. ~Bozhevolnyi,
\enquote{Theoretical analysis of square surface plasmon-polaritons
waveguides for long-range polarization-independent waveguiding,}
Phys. Rev. B,
  \textbf{76}, 035434 (2007).

\bibitem{Palikbook}
E.~D. Palik, \emph{Handbook of Optical Constants of Solids}
(Academic, New
  York, 1985).

\bibitem{Hausbook}
H.~A. Haus, \emph{Waves and fields in optoelectronics}
(Prentice-Hall,
  Englewood Cliffs, N. J., 1984).

\bibitem{Johnson:1998}
S.~G. Johnson, C.~Manolatou, S.~H. Fan, P.~R. Villeneuve, J.~D.
Joannopoulos,
  and H.~A. Haus, \enquote{Elimination of cross talk in waveguide
  intersections,} Opt. Lett. \textbf{23}, 1855--1857 (1998).

\bibitem{Maier:2006}
S. A. ~Marier, \enquote{Gain-assisted propagation of
electromagnetic energy in subwavelength surface plasmon polariton
gap waveguides,} Opt. Commun. \textbf{258}, 295--299 (2006).

\bibitem{Nezhad:2006}
M. P. ~Nezhad. K. Tetz, and Y. Fainman, \enquote{Gain assisted
propagation of surface plasmon polaritons on planar metallic
waveguides,} Opt. Express \textbf{12}, 4072--4079 (2006).


\end{thebibliography}
\end{document}